\documentclass{ieeeaccess}
\usepackage{cite}
\usepackage{amsmath,amssymb,amsfonts}

\usepackage{graphicx}
\usepackage{textcomp}
\usepackage{caption}
\usepackage{algorithm}
\usepackage{algorithm}
\usepackage{algpseudocode}  
\usepackage{bm}
\usepackage{algorithmicx}

\def\BibTeX{{\rm B\kern-.05em{\sc i\kern-.025em b}\kern-.08em
		T\kern-.1667em\lower.7ex\hbox{E}\kern-.125emX}}
\begin{document}
	\history{Date of publication xxxx 00, 0000, date of current version xxxx 00, 0000.}
	\doi{10.1109/ACCESS.2017.DOI}
	
	\title{SSNE: Effective Node Representation for Link Prediction in Sparse Networks}
	\author{\uppercase{Ming-Ren Chen}\authorrefmark{1}, \uppercase{Ping Huang}\authorrefmark{1}, \uppercase{Yu Lin}\authorrefmark{4}, \uppercase{Shi-Min Cai}\authorrefmark{1,2,3}}
	\address[1]{ School of Computer Science and Engineering, University of Electronic Science and Technology of China, Chengdu 611731, China}
	\address[2]{Institute of Fundamental and Frontier Sciences, University of Electronic Science and Technology of China, Chengdu 611731, China}
	\address[3]{Big data research center, University of Electronic Science and Technology of China,  Chengdu 611731, China}
	\address[4]{ Research School of Computer Science, Australian National University, Canberra, Australian}
	
	\tfootnote{This work is supported by National Natural Science Foundation of China (No. 61673086) and the Science Promotion Programme of UESTC, China (No. Y03111023901014006).}
	
	\markboth
	{Ming-Ren Chen \headeretal: SSNE: Effective Node Representation for Link Prediction in Sparse Networks}
	{Ming-Ren Chen \headeretal: SSNE: Effective Node Representation for Link Prediction in Sparse Networks}
	
	\corresp{Corresponding author: Shi-Min Cai (e-mail: shimin.cai81@gmail.com).}

\begin{abstract}
Graph embedding is gaining popularity for link prediction in complex networks. However, few works focus on the effectiveness of graph embedding models on link prediction in sparse networks. This paper proposes a novel graph embedding model, \textbf{S}parse \textbf{S}tructural \textbf{N}etwork \textbf{E}mbedding (SSNE), to obtain node representation for link predication in sparse networks. The SSNE first transforms the adjacency matrix into the \textbf{S}um of \textbf{N}ormalized \textbf{H}-order \textbf{A}djacency \textbf{M}atrix (SNHAM) and then maps the SNHAM matrix into a $d$-dimensional feature matrix for node representation via a neural network model. The mapping operation is proved to be an equivalent variety of singular value decomposition. Finally, we calculate nodal similarities for link prediction based on the $d$-dimensional feature matrix. The extensive testing experiments based on artificial and real sparse networks suggest that the SSNE shows the effective node representation for link prediction in sparse networks, supported by the better link prediction performance compared to those of structural similarity indexes, matrix optimization, and other graph embedding models.
\end{abstract}


\titlepgskip=-15pt

\maketitle

\section{Introduction}
\label{sec1}
\PARstart{I}{N} natural complex systems, there are many entities, which interact with each other in a complicated way. By treating these 
entities as nodes and the corresponding interactive relationships as edges, we can abstract such systems into the network (or graph)
model. Naturally, diverse types of complex networks are available to represent real complex systems, such as social networks,  traffic networks, brain and biological networks, infrastructure networks, etc. \cite{boccaletti2006complex,bullmore2009complex,boccaletti2014structure}. 
Complex networks are continually evolving, and new connections between entities may occur in the future. Therefore, link prediction becomes an important task to study network structure's dynamic evolution \cite{getoor2005link,liben2007link,lichtenwalter2010new,lu2011link,martinez2017survey,haghani2019systemic}.  


In previous researches, a relatively simple link prediction framework is proposed based on the assumption that the greater the similarity between two nodes in the network, the greater the possibility of a connection between them \cite{liben2007link}. Then, many similarity measurements of nodes have been proposed to compute similarity-based indexes for link prediction. A network contains a massive amount of structural information, which has been modeled as many similarity-based indexes, including the common neighbor (CN) index \cite{lorrain1971structural}, the Adamic-Adar (AA) index \cite{adamic2003friends}, the resource allocation (RA) index \cite{zhou2009predicting}, the Katz index \cite{katz1953new}, the restarted random walk (RWR) index \cite{brin1998anatomy}, and the SimRank index \cite{jeh2002simrank}, etc. These indexes can mainly be divided into two categories, local and global structural similarity indexes. The local structural similarity indexes (e.g., CN, AA, RA) only use the local topological information of nodes, which benefit from low computational complexity and become suitable for large-scale networks. However, their accuracy is slightly lower compared to that of the global structural similarity indexes (e.g., Katz, RWR, and SimRank), which considers the global topological information at a higher computational cost.

Graph embedding (i.e., graph representation) has been widely used in link prediction problems with representation learning development\cite{xiao2015one,cao2019network}. Graph embedding can map a graph into low-dimension vector space, and at the same time, keep the structure feature and inherent attribute of the graph \cite{luo2017local,gu2018rare,ni2018co,nguyen2018continuous,shi2019unsupervised}. Commonly, its pivotal is to sample enough structural information by random walks on a graph (or network). For example, DeepWalk \cite{perozzi2014deepwalk} is one of the most popular random-walk-based graph embedding models. The link prediction method based on DeepWalk is shown to predict better the possible incidence of MicroRNA genetic disease \cite{li2017predicting,chen2019predicting}, as well as individual multiple interests or attributes \cite{jin2016predicting,yu2020structured}. Although these embedding models succeed in link prediction in many natural networks, they involve critical experience-tuned parameters, such as the sampling length of a random walk and the number of random walks \cite{perozzi2014deepwalk}. A typical scenario may only specify a locally maximum value within a finite interval of experience-tuned parameters. The error accumulation of multiple parameters would also hinder link prediction performance in sparse networks.

Therefore, in the framework of graph embedding, we propose a novel graph embedding model, \textbf{S}parse \textbf{S}tructure \textbf{N}etwork \textbf{E}mbedding (\textbf{SSNE}), to solve the problems mentioned above of random-walk-based graph embedding models. The SSNE includes two essential operations. The first is to transform the adjacency matrix of a sparse network into a general mutual information matrix based on the algorithm of the \textbf{S}um of \textbf{N}ormalized \textbf{H}-order \textbf{A}djacency \textbf{M}atrix (\textbf{SNHAM}), and the second is to map the SNHAM matrix into a $d$-dimensional feature matrix for effective node representation via a neural network model. The details will be introduced in \textbf{Section \uppercase\expandafter{\romannumeral4}}. In further, we design experiments over various datasets to verify the effectiveness of SSNE for link prediction. The experimental results based on sparse networks show that the link prediction method based on SSNE outperforms other methods based on structural similarity indexes, matrix optimization, and other graph embedding models. As for relatively dense and better-structured networks, SSNE still shows comparable performance as structural similarity indexes, which is better than matrix optimization.

In short, in this paper, we make the following contributions:
\begin{itemize}
\item We propose a novel graph embedding model that overcomes the drawbacks in the prevail random-walk-based graph embedding models. The SNHAM algorithm is used to transform the adjacency matrix into a new matrix of theoretical co-occurrence probability between each pair of nodes, which substitutes the sampling method in random-walk-based graph embedding models. Meanwhile, we testify the mapping operation of the SNHAM algorithm to be an equivalent variation of the singular value decomposition (SVD), which significantly improves the computational efficiency of obtaining the feature matrix.
\item We construct a link prediction method based on SSNE. The testing experiments' results based on six real networks and two types of artificial network models show the excellent performance of link prediction in sparse networks.
\item We verify the algorithmic stability of link prediction method based SSNE by selecting different combinations of parameters. The results show that the proposed method is generally insensitive to parameters.
\end{itemize}

The remainder of this paper is organized as follows. In \textbf{Section \uppercase\expandafter{\romannumeral2}}, we briefly survey related work. \textbf{Section \uppercase\expandafter{\romannumeral3}} gives the problem definition. \textbf{Section \uppercase\expandafter{\romannumeral4}} presents the whole framework of the link prediction method based on SSNE in detail. \textbf{Section \uppercase\expandafter{\romannumeral5}} contains experimental material and evaluation. \textbf{Section \uppercase\expandafter{\romannumeral6}} presents the experimental result and discusses the effectiveness of adjustable parameters in link prediction performance. We finally conclude the paper in \textbf{Section \uppercase\expandafter{\romannumeral7}}.

\section{Related works}
\label{sec2}

This section briefly illustrates the related works in two aspects. On the one hand, we introduce some classical link prediction methods based on structural similarity index and discuss corresponding research achievements in recent years. On the other hand, we also discuss some popular graph embedding models based on representation learning for link prediction in complex networks.

\subsection{Link prediction based on structural similarity index}
\label{subsec1}
The structural similarity indexes are defined by the similarity between two nodes based on their corresponding local or global structural measurements. The common structural similarity indexes include the CN index, the AA index, the RA index, the Katz index, the RWR index, the SimRank index, \emph{etc}. (refer to \cite{lu2011link} for more indexes). The CN index calculates the similarity of a pair of nodes by counting their number of common neighbors. The AA index \cite{adamic2003friends} and RA index \cite{zhou2009predicting}, based on the CN index, punish the common neighbors with large degrees respectively by the inverse logarithm and the reciprocal of common neighbors' degrees. The Katz index \cite{katz1953new} can distinguish different neighboring nodes' influences and assign different weights to the neighboring nodes, causing the weights to be biased towards short paths. The RWR index is an alternative method of the PageRank algorithm and is defined by the probability of random walk between a pair of nodes in the steady-state \cite{brin1998anatomy}. The SimRank index \cite{jeh2002simrank} also involves the random walk process and measures how soon two random walkers, respectively starting from the endpoints, are expected to meet a certain node. These structural similarity indexes have been widely used to infer linking probability for link prediction in complex networks. Herein, we illustrate some very recent works on the link prediction methods based on structural similarity indexes.

Inspired by the above common structural similarity indexes, sever recent works synthesized more structural measurements to form complicated structural similarity indexes for link prediction. In \cite{zhou2018h}, Zhou \emph{et al.} replaced the degree with H index to form H-index-based link prediction methods, which significantly improve link prediction accuracy. In \cite{zhu2018roles,zhu2018hybrid}, Zhu et al. discussed the roles of degree, H-index, and coreness in link prediction in complex networks and formed a hybrid similarity index synthesizing degree and H-index. In \cite{pech2019link}, Pech et al. proposed a simple assumption that the likelihood of the existence of a link between two nodes can be unfolded by a linear summation of neighboring nodes’ contributions and obtained the optimal likelihood matrix that shows remarkably better-predicting prediction performance.

\subsection{Link prediction based on graph embedding}
\label{subsec2}
Graph embedding is used to map network structure into low-dimensional vector space indicated by a feature matrix of nodes reduced from an adjacency matrix. Based on the nodes' feature matrix, the similarity index is defined by the similarity between feature vectors of nodes. Thus, the link prediction method based on graph embedding strongly correlates with the graph embedding models. Herein, we illustrate some previous works on embedding graph models.

In recent years, graph embedding models have attracted more attention. In \cite{perozzi2014deepwalk}, Perozzi \emph{et al.} proposed DeepWalk, where the random walk sampling processes produce linear sequences of nodes, and these sequences are used to calculate co-occurrence probabilistic matrix of nodes that are mapped into a feature matrix by a neural network model. In \cite{tang2015line}, Tang et al. explicitly defined two objective functions, 1st-order proximity and 2nd-order proximity, to obtain the topological information of network structure.
They then used the linear combination of such proximity to represent the global proximity. In \cite{grover2016node2vec}, Grover et al. proposed Node2Vec, which maintained the high order proximity between nodes by maximizing the probability of subsequent nodes in the random traversal graph. Compared with DeepWalk, Node2Vec has made some improvements in the random walk sampling process.

\section{Problem Definition}
\subsection{Pre-Definition}

An undirected unweighted network is represented by $G= \langle V, E \rangle$ where the node set V=\{\ {$v_{1}$,$v_{2}$,...,$v_{n}$} \}\ and the edge set 
$E=\left\{e_{i, j}\right\}(i, j \in V)$. The dynamic evolution of network structure is represented by multiple snapshots of 
network, i.e., $\mathbb{G}=\left\{G_{1}, G_{2}, \dots, G_{t-1}, G_{t}, \dots, G_{N-1}, G_{N}\right\}$. At the current time $t$, $G_{t}=<V_{t}, E_{t}>$ denotes 
a snapshot of the network. Naturally, $G_{t-1}=<V_{t-1}, E_{t-1}>$ shows a previous snapshot of the network. \t{We assume that} the node set is stable, \emph{i.e.,} $V_{1}=V_{2}=\cdots=V_{t-1}=V_{t}$, 
but the edge set is dynamically changing, which characterizes the dynamic evolution of network structure. 

For simplicity, any two different nodes are indicated by symbols $u$ and $v$, and the adjacency matrix of the network is expressed by symbol $A$. Obviously, 
if there exists an edge between nodes $u$ and $v$, $A(u, v)=1$, otherwise $A(u, v)=0$. For a node $u$, its adjacency vector is $A_{u}^{1 \times n}=A(u,:)$. 
We assume that the feature matrix $R$ for node representations is obtained from the dimensionality reduction of the adjacency matrix. 
In a similar way, for a node $u$, its $d$-dimensionality vector is $R_{u}^{1 \times d}=R(u,:)$.

We illustrate the important symbols involved in the model of SSNE. In the SNHAM algorithm, the output is defined by matrix $SNHAM \in \mathbb{R}^{n \times n}$, and the specific order is set as $h$. 
Because the elements of SNHAM matrix can reflect the co-occurrence probability for each pair of nodes, a node $u$ has $n$-dimension vector of  co-occurrence probability, $SNHAM_{u}^{1 \times n}= SNHAM (u,:)$.
In the single-hidden layer feedforward neural network model, the input is defined by the matrix $X \in \mathbb{R}^{n \times n}$, 
and the kernel and activation function between the input and hidden layers is respectively set as  $W_{1}$ and $f_{1}(x)$; the output is defined by the matrix $Y \in \mathbb{R}^{n \times n}$, and the kernel and activation function between the hidden and output layers is respectively set as  $W_{2}$ and $f_{2}(x)$; the feature matrix $R \in \mathbb{R}^{n \times d}$ of node representations is obtained in the hidden layer. The more details of symbol description are summarized in \textbf{Table 1}.

\begin{table}
  \caption{Notation note of each symbol}
  \label{Table:1}
  \centering
  \begin{tabular}{|c|c|}
    \hline
    Symbol&Comments\\
    \hline
    $u$ or $v$ & denote a node\\
    $n$ & the number of nodes, i.e., the full dimension\\
    $d$ & the target of dimensionality reduction\\
    $h$ & the orders of SNHAM algorithm\\
    $t$ & the current time\\
    $t-1$ & the previous time\\
    $G_{t}$ & a snapshot of network at time $t$\\
    $V_{t}$ & the node set in network  $G_{t}$\\
    $E_{t}$ & the edge set in the network $G_{t}$\\
    $W_{i}$ & the kernel of $i$th-layer network \\
    $f_{i}(x)$ & the activation function of $i$th-layer network\\
    $\mathbb{G}$ & multiple snapshots of network\\
    $A$ & the adjacency matrix of network\\
    $A_{u}^{1 \times n}$ & the adjacency vector for node $u$\\
    $SNHAM$ & the output matrix of the SNHAM algorithm\\
    $SNHAM_{u}^{1 \times n}$ & the SNHAM vector for node $u$\\
    $R$& the feature matrix of node representations\\
    $R_{u}^{1 \times d}$ & the $d$-dimension vector for node $u$\\
    $X$ & the input matrix of neural network model\\
    $X_{u}^{1 \times n}$ & the input vector for node $u$\\
    $Y$ & the output matrix of neural network model \\
    $Y_{u}^{1 \times n}$ & the output vector for node $u$\\
  \hline
\end{tabular}
\end{table}

\subsection{Problem Statement}
Studying the whole dynamic evolution of network $\mathbb{G}$ is a complicated and challenging task. In order to simplify the process of the derivation, we herein only consider the relationship between the current $t$ and previous time $t-1$, that is $\mathbb{G}=\left\{G_{t-1}, G_{t}\right\}$. Therefore, inferring dynamic evolution of network structure from $t-1$ to $t$ realized by the link prediction based on $G_{t-1}$ and $G_t$. The training set and test set can be set by $G_{t-1}$ and $G_{t}-G_{t-1}$, respectively. Note that the real (benchmark) networks in the testing experiments aren't temporal (i.e., absent of the time scale). We thus assume the original network as $G_t$ and hide a proportion of its edges to assume the residual network as $G_{t-1}$. Based on $G_{t-1}$, our task is to get the feature matrix of node representations that meets the lower dimension. Still, it involves a large number of topological information of network structure and then applies the feature matrix to predict the hidden edges.

\section{SSNE for link prediction}

In this section, we describe the model of SSNE in detail. As shown in \textbf{Figure 1}, the SSNE consists of two steps. First, we introduce the SNHAM algorithm to obtain its corresponding matrix $SNHAM$ that can reflect the theoretical values of co-occurrence probability between each pair of nodes.
Then, we design a neural network model to calculate the corresponding co-occurrence probability (i.e., the output matrix $Y$). According to the difference between the matrices $SNHAM$ and $Y$, the loss function is established. Using the stochastic gradient descent approach to minimize the loss function, we can get the optimal kernels and determine the feature matrix $R$ in the hidden layer. However, the stochastic gradient descent approach has high time complexity in its iterative operation. We then find an alternative method to directly map the $\log (\mathrm{SNHAM})$ matrix into the feature matrix of node representations and demonstrate that the mapping operation is an equivalent variation of SVD. Finally, we apply the results of the SSNE to construct the similarity index for link prediction.

\begin{figure*}[ht]
  \centering
  \includegraphics[width=0.8\linewidth]{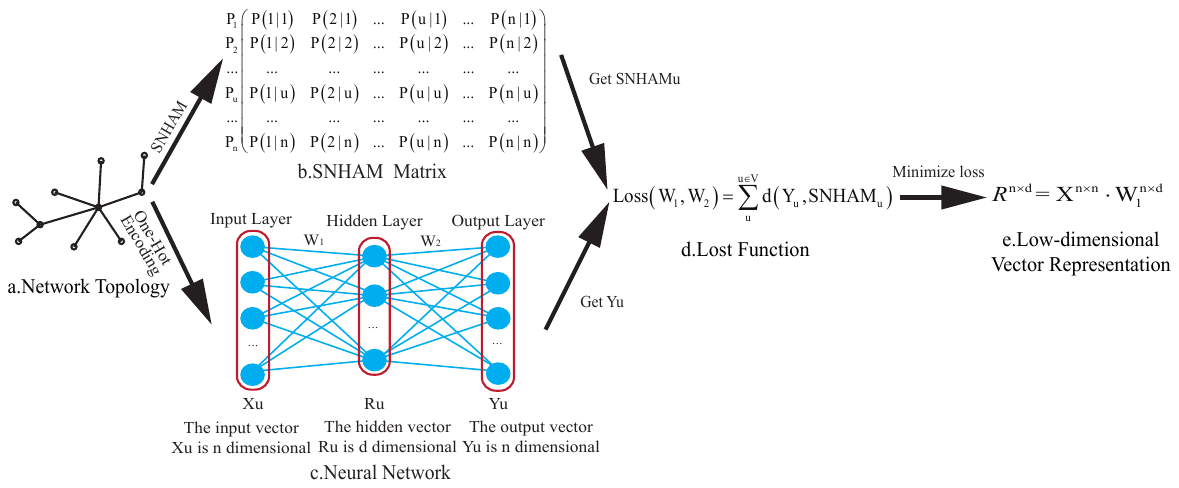}
  \caption{(Color online) The schematic description of the SSNE. (a) An example of network topology; (b) The construction of SNHAM matrix; (c) The neural network model that is used to acquires the low-dimensional feature matrix; (d) The loss function of neural network model; (e) The acquirement of low-dimensional vector representation by minimizing the loss function.}
\end{figure*}

\subsection{SNHAM matirx}

Although using the random walk sampling process is effective in converting the topological information of network structure into linear sequences, it has been found that this method has obvious drawbacks. As mentioned above, graph embedding models based on random walk need to determine some random walk parameters, such as the sampling length of a random walk and the number of random walks, so that they are sensitive to such parameters. More importantly, we can only determine the random walk's empirically optimal parameters (i.e., local best at a finite parameter interval). Further, the finite-length linear sequences collected by the random walks have vital errors in representing the boundary nodes. Thus, the multiple parameters' accumulative errors significantly affect the accuracy of link prediction in complex networks.
To solve the above problem existing in the graph embedding models based on a random walk, we propose the SNHAM algorithm to capture the network structure's topological information directly. We label the nodes in the network and order them to obtain an adjacency matrix $A$. First, we normalize the adjacency matrix by row to get the 1st-order transition probability matrix $S_{1}$. The row normalization function is set as
$\mathrm{Normal}(X)$, so the above operation can be expressed by

\begin{equation}
S_{1}=\mathrm{Normal}(A) .
\end{equation}

In a similar way, we calculate the $h$-order transition probability matrix $S_{h}$ by the $h$-order reachable matrix $A^{h}$ ($h=1,2,\cdots, h$),

\begin{equation}
S_{h}=\mathrm{Normal}\left(A^{h}\right), \text{where} \ A^{h}=\overbrace{A \times A \cdots \times A}^{h}.
\end{equation}

Then, we define the $h$-order similar probabilistic co-occurrence matrix $SPCO_{h}$, which is the sum of probability transition moments considering a restart. The restart probability is set as $\alpha$. Thus, $SPCO_{h}$ is described by
\begin{equation}
SPCO_{h}=\sum_{i=1}^{h}\left((1-\alpha) S_{i}+\alpha S_{1}\right).
\end{equation}

We consider that the restart process is excluded (i.e., $\alpha=0$), that is, $SPCO_{h}$ can be reduced to the following form,

\begin{equation}
SPCO_{h}=\sum_{i=1}^{h} S_{i}.
\end{equation}

Finally, we normalize the rows of $SPCO_{h}$ matrix, and the final result is denoted as the SNHAM matrix, which can be expressed as follows:

\begin{equation}
SNHAM=\mathrm{Normal}\left(SPCO_{h}\right).
\end{equation}

The SNHAM algorithm can efficiently obtain the locally topological information of network structure and effectively solve the random walk sampling process's drawbacks. As the restart process is excluded, the adjusting parameter in the SNHAM algorithm is only the order $h$. The single parameter can avoid the random walk sampling process's accumulative errors of multiple parameters.
Simultaneously, the SNHAM algorithm to obtain the network structure is no longer transforming the network structure into linear sequences of nodes. There don't exist errors in the process of representing the boundary node of each linear sequence. We show the pseudocode of the SNHAM algorithm in \textbf{ Algorithm 1}.


\begin{algorithm}[ht] 
	\caption{SNHAM}  
	\begin{algorithmic}[1]  
	\Require    adjacency matrix \bm{$A$} of $G_{t-1}$; order index \bm{$h$} 
	\Ensure  SNHAM matrix \bm{$SNHAM$}
	\State	Initializing an $n \times n$  matrix $SPCO$; 
	\For{each $i \in [1, h]$}  
		\State Calculating $A^{i}=\overbrace{A \times A \cdots \times A}^{i}$;  
		\State Normalizing matrix $A^{i}$ by row, $S_{i}=$ Normal $\left(A^{i}\right)$; 
		\State   $SPCO_h$=$\sum_{i=1}^{h}$ $S_{i}$
    \EndFor  
    \State Normalizing matrix $SPCO_h$ by row, SNHAM= Normal $(SPCO_h)$;
  \end{algorithmic}  
\end{algorithm} 

\subsection{Neural network model}

The neural network is widely used to study multi-level feature representation, and the results obtained from representation learning are proved to be successfully applied in various fields. Herein, we use the single-hidden layer feedforward neural network model to construct high-quality and low-dimensional feature representation based on the SNHAM matrix. It is assumed to be a potential nonlinear mapping relationship between the vector representation space of the SNHAM matrix and the low-dimensional feature representation space. In SSNE, the single-hidden layer feedforward neural network model based on the SNHAM matrix is designed to calculate the co-occurrence probability matrix calculation.

Specifically, for a given node $u$, we input its one-hot coding vector $X_{u}^{1 \times n}$, and map $X_{u}^{1 \times n}$ into a low-dimensional vector $R_{u}^{1 \times d}$ through the kernel $W_{1}$ and activation function $f_{1}(X)=\mathrm{X}$,

\begin{equation}
R_{u}^{1 \times d}=f_{1}\left(X_{u}^{1 \times n} \cdot W_{1}^{n \times d}\right)=X_{u}^{1 \times n} \cdot W_{1}^{n \times d}.
\end{equation}

Referring to the neural network model, the low-dimensional vector $R_{u}^{1 \times d}$ is able to be mapped into the co-occurrence probabilistic vector $Y_{u}^{1 \times n}$ through the kernel $W_{2}$ and activation function $f_{2}(X)=\operatorname{Softmax}(\mathrm{X})=\frac{e^{x}}{\Sigma_{i} e^{i}}$,

\begin{equation}
Y_{u}^{1 \times n}=\operatorname{Softmax}\left(R_{u}^{1 \times d} \cdot W_{2}^{d \times n}\right).
\end{equation}

We use the theoretical co-occurrence probabilistic vector $SNHAM_{u}^{1 \times n}$ of node $u$ obtained from the SNHAM matrix and compare it with $Y_{u}^{1 \times n}$ via the Euclid measurement. The loss function
${L}\left(W_{1}, W_{2}\right)$ is built by summing the errors across over all nodes,

\begin{equation}
{L}\left(W_{1}, W_{2}\right)=\sum_{u \in V} \mathrm{d}\left(SNHAM_{u}, \operatorname{Softmax}\left(\left(X_{u} \cdot W_{1}\right) \cdot W_{2}\right)\right).
\end{equation}

The kernels $W_{1}$ and $W_{2}$ are obtained through the stochastic gradient descent approach by minimizing the loss function. We focus the low-dimensional feature matrix in the hidden layer, which is described by

\begin{equation}
\text R^{n \times d}=X^{n \times n} \cdot W_{1}^{n \times d}.
\label{e9}
\end{equation}

As the stochastic gradient descent approach is high computational complexity, we provide an alternative method in the following subsection to improve the computational efficiency of obtaining a feature matrix significantly.

\subsection{Minimizing $L(W_1,W_2)$ by SVD}

The above-mentioned optimization procedure of minimizing the loss function ${L}\left(W_{1}, W_{2}\right)$ is actually equivalent to make $\operatorname{Softmax}\left(\left(X \cdot W_{1}\right) \cdot W_{2}\right)$ approximate $SNHAM$ by adjusting the kernels $W_{1}$ and $W_{2}$. An ideal situation is that ${L}\left(W_{1}, W_{2}\right)=0$, which satisfies the condition,

\begin{equation}
SNHAM_{i}=\operatorname{Softmax}\left(\left(X_{i} \cdot W_{1}\right) \cdot W_{2}\right), \text{where} \ i = 1, 2, \cdots, n.
\label{e10}
\end {equation}

We further simplify the variable $\left({X_{i}} \cdot W_{1}\right) \cdot W_{2}$. Since the input matrix $X$ encoded for the one-hot form is actually an identity matrix, we can write $W_{1} \cdot W_{2}$ as the product matrix $Z$. Then, equation (\ref{e10}) can be rewritten as

\begin {equation}
SNHAM_{i}=\operatorname{Softmax}\left(Z_i \right), \text{where} \ i = 1, 2, \cdots, n.
\label{e11}
\end {equation}

Supposing equation (\ref{e11}) has an inverse function, $Z_{i}$ can be written as

\begin {equation}
 Z_{i}=\operatorname{Softmax}^{-1}\left(SNHAM_{i}\right).
\label{e12}
\end {equation}

Naturally, the main task turns to obtain such inverse function. We set a input vector $x_{i}=\left(x_{i, 1}, x_{i, 2}, \ldots, x_{i, j}, \ldots, x_{i, n}\right)$, and the output vector via Softmax function is denoted as $y_{i}=\left(y_{i, 1}, y_{i, 2}, \ldots, y_{i, j}, \ldots, y_{i, n}\right)$. Without loss of generality, each value $x_{i,j}$ producing a corresponding $y_{i,j}$ satisfies an equation,

\begin {equation}
{y}_{i, j}=\operatorname{Softmax}\left(x_{i, j}\right)=\frac{e^{x_{i, j}}}{\sum e^{x_{i, j}}}.
\label{e13}
\end {equation}
When the input vector is determined, $\sum e^{x_{i, j}}$ is a constant that is set as $k_{i}$. The conditions are satisfied with

\begin {equation}
\text { s.t. } \quad\left\{\begin{array}{c}{\sum_{j=1}^{n} y_{i, j}=1} \\ \\ {\sum_{j=1}^{n} e^{x_{i, j}}=k_{i}}\end{array}\right. ,
\label{c13}
\end {equation}
and used to obtain a variation of equation (\ref{e13}), then we can get the following formula,

\begin {equation}
x_{i, j}=\log \left(k_{i} \cdot y_{i, j}\right).
\label{e14}
\end {equation}

Inspired by equation (\ref{e14}), we assume the inverse function with a formula,

\begin {equation}
\operatorname{Softmax}^{-1}\left(y_{i, j}\right)=x_{i, j}=\log \left(c_{i} \cdot y_{i, j}\right).
\label{e15}
\end {equation}

For a certain $x_i$, equation (\ref{e15}) is determined only when $c_i$ is constant. In further, we verify the above-mentioned assumption. Equation (\ref{e15}) is generalized as

\begin {equation}
\left\{\begin{aligned} x_{i, 1} &=\log \left(c_{i} \cdot y_{i, 1}\right) \\ x_{i, 2} &=\log \left(c_{i} \cdot y_{i, 2}\right) \\ ......\\ x_{i, n} &=\log \left(c_{i} * y_{i, n}\right) \end{aligned}\right. ,
\end {equation}
which is equivalent to the following formula,

\begin {equation}
\left\{\begin{aligned} e^{x_{i, 1}} &=c_{i} \cdot y_{i, 1} \\ e^{x_{i, 2}} &=c_{i} \cdot y_{i, 2} \\......\\ e^{x_{i, n}} &=c_{i} \cdot y_{i, n} \end{aligned}\right. ,
\label{e17}
\end {equation}

We sum the left and right terms in equation (\ref{e17}) and obtain the following formula,

\begin {equation}
\sum e^{x_{i, j}}=c_{i} \cdot \sum y_{i, j}.
\label{e18}
\end {equation}

According to the conditions in equation (\ref{c13}), we obtain $c_i=k_i$ from equation (\ref{e18}), which implies that for a certain $x_i$, $c_i$ is a constant. Thus, the specific formula of the inverse function is independent of $c_i$. To make it easy to calculate, we set all $k_{i}$ to 1 by assuming the independence of input vectors so that the inverse function is simplified as

\begin {equation}
\text { Softmax }^{-1}\left(y_{i, j}\right)=x_{i, j}=\log \left(y_{i, j}\right).
\label{e19}
\end {equation}

Turning to equation (\ref{e12}), it is specified as

\begin {equation}
Z_{i}=\operatorname{Softmax}^{-1}\left({SNHAM}_{i}\right)=\log \left({SNHAM}_{i}\right).
\label{e20}
\end {equation}

Considering the zero value of co-occurrence probability in the SNHAM matrix, we uniformly add a very small positive $\sigma$ ($\sigma=10^{-8}$ in the testing experiments). We finally obtain the inverse function with the formula,

\begin {equation}
Z=\log (SNHAM+\sigma)=\log (SNHAM^{\prime}).
\label{e21}
\end {equation}
Through equation (\ref{e21}), the specific matrix $Z$ is also acquired.

We have known $Z=W_{1} \cdot W_{2}$, and divide the matrix $Z$ by SVD to get $W_{1}$, $W_{2}$ easily. The SVD procedure of $\log (SNHAM^{\prime})$ is approximately equivalent to the optimization procedure of the neural network model. Without loss of generality, we denote the decomposition
process as

\begin {equation}
\log (SNHAM^{\prime})=U \Sigma V^{T} .
\end {equation}

We choose the first $d$ largest singular values , and approach $log(SNHAM^{\prime})$ to $log(SNHAM^{\prime})_{d}$ , according to the following formula,

\begin {equation}
\log ({SNHAM^{\prime}}) \approx \log ({SNHAM^{\prime}})_{d}=U_{d} \Sigma_{d} V_{d}^{T}
\label{e23}
\end{equation}

According to equation (\ref{e23}), we easily obtain $W_1=U_{d} \Sigma_{d}$ and $W_2=V_{d}^{T}$. Finally, according to equation (\ref{e9}),
the $d$-dimensional feature matrix $R$ can be expressed as

\begin {equation}
\text R^{n \times d}=U_{d} \Sigma_{d}.
\end{equation}

\subsection{Similarity index based on feature matrix}

After the original network topology is represented by the $d$-dimension feature matrix by the SSNE, we use such a feature matrix to construct a similarity index for link prediction.
For any unknown edge $e_{u,v}$ between a pair of nodes $u$ and $v$, its potential probability is quantified by these two nodes' similarity index. The similarity measurement is used by the Euclidean distance between the feature vectors of $u$ and $v$, which is described as

\begin {equation}
D(e_{u,v})=\sqrt{\left(x_{1v} -x_{1u} \right)^{2}+\left(x_{2v} -x_{2u}\right)^{2}+\cdots+\left(x_{dv} -x_{du} \right)^{2}}
\end{equation}

Considering the inverse correlation that the greater the distance is, the lower the similarity is, we take its reciprocal and add 1 to $D(e_{u,v})$ to prevent the case that $D(e_{u,v})$ is zero or too small. Finally, the similarity index is constructed by

\begin {equation}
S(e_{u,v})=\frac{1}{1+D(e_{u,v})}.
\end {equation}

In the link prediction in complex networks, the higher the similarity index, the higher the potential possibility the unknown edge will be linked. We show the link prediction method's pseudocode based on SSNE in \textbf{ Algorithm 2}.


\begin{algorithm}[ht] 
  \caption{Link prediction based on SSNE}  
  \begin{algorithmic}[1]  
	\Require  SNHAM Matrix \bm{$SNHAM$}; dimension \bm{$d$} 
	\Ensure  Evaluation Index \bm{$AUC$}
	\State Calculating $\log (\mathrm{SNAHM}+\sigma)$;
	\State $\mathrm{U} \Sigma V^{T}=SVD(\log (S P M I+\sigma))$
	\State Choosing $d$ largest singular value, $\mathrm{U}_{d} \Sigma_{d} V_{d}^{T}\approx \mathrm{U} \Sigma V^{T}$
	\State Obtaining feature matrix R, $R^{n \times d}=U_{d} \Sigma_{d}$
	\State Calculating Euclidean distance, \\ $D(e_{u,v})=\sqrt{\left(x_{1v} -x_{1u} \right)^{2}+\cdots+\left(x_{dv} -x_{du} \right)^{2}}$
	\State Calculating similarity index, $S(e_{u,v})=\frac{1}{1+D(e_{u,v})}$
	\State Initializing sampling parameter of AUC, $N=672400$, $N^{\prime}=0$, $N^{\prime\prime}=0$;
	\For{each $ i \in [1,N]$}  
		\If{$S_{i}\left(e_{a, b}\right)>S_{i}\left(e_{c, d}\right)$}{\ $N^{\prime}+=1$}
		\ElsIf{$S_{i}\left(e_{a, b}\right) = S_{i}\left(e_{c, d}\right)$}{\ $N^{\prime\prime}+=1$}
		\Else{\ other cases}
		\EndIf
		\EndFor  
	\State Calculating AUC, $AUC=\frac{N^{\prime}+0.5 \cdot N^{\prime\prime}}{672400}$
	\end{algorithmic}  
\end{algorithm} 

\section{Experimental material and evaluation}

We design testing experiments based on six real networks and two types of artificial network models to validate the effectiveness of SSNE for link prediction in complex networks. In this section, the specific descriptions of real networks, two types of artificial networks, and the evaluation are illustrated, respectively.

\subsection{Real networks}
We show six real networks that are described as:

\textbf{Brain}\cite{konect}: It is the neuronal connection network of a rhesus macaque. The nodes and edges represent neurons and fiber bundles among these neurons, respectively. In this network, there are 242 neurons, 3,054 fiber bundles, and the average degree of the network is 25.24.

\textbf{Yeast}\cite{batagelj2009pajek}: It is the biological network in budding yeast. The nodes and edges represent proteins and interactions among these proteins. There are 2,375 proteins and 11,693 known interactions in this network, and the average degree of network is 9.85.

\textbf{Air}\cite{konect}: It is the traffic network of air control. The nodes and edges represent airports or service centers and the preferred air route among these airports or service centers recommended by the NFDC (National Flight Data Center). In this network, there are 1,226 airports or services centers, 2,410 preferred air routes, and the average degree of the network is 3.93

\textbf{Road}\cite{konect}: It is the road network in Minnesota state. The nodes and edge represent voluntary buildings and the direct road between these buildings. In this network, there are 2,642 buildings, and there are 3,303 direct roads, and the average degree of the network is 2.50.

\textbf{Twitter}\cite{rossi2012fastclique}: It is the forwarding network of Twitter users about Obama's re-election as President of the United States in 2012. The nodes and edges represent twitter users and retweeting relationships between these users, respectively. There are 3,212 Twitter users in this network, 3,423 retweeting connections, and the network's average degree is 2.13.

\textbf{Power}\cite{konect}: It is the west power grid in the U.S. The nodes and edges represent substations or converters and high-voltage lines among these substations or converters. There are 4,941 substations or converters in this network in this network, 6,594 high-voltage lines, and the average degree of the network is 2.70.

We also summarize the basic topological information of six real networks, including the number of nodes and edges, the edge sparsity, the average degree, the clustering coefficient, and the degree heterogeneity, which are shown in
\textbf{Table 2}.

\begin{table}
  \caption{Basic topological information of six real networks. $|V|$ and $|E|$ indicate the number of nodes and edges,  $ \langle k \rangle$ 
   is the average degree, $ES$ is the edge sparsity, $ \langle d \rangle$ is the average distance, $C$ is the clustering coefficient, and $H=\frac{\langle k^2 \rangle}{{\langle k \rangle}^2}$ is the degree heterogeneity.}
  \label{Table:2}
  \centering
   \resizebox{\linewidth}{!}{
  \begin{tabular}{c|c|c|c|c|c|c|c}
    \hline  \hline
    Nets & $|V|$ & $|E|$ & $\langle k \rangle$& $ES$ & $\langle d \rangle$ & $C$ & $H$ \\
    \hline
    $Brain$ & 242 & 3,054 & 25.24 & 0.1047& 2.22 & 0.450 & 1.53 \\
    $Yeast$ & 2,375 & 11,693 & 9.85 & 0.0041 & 5.09 & 0.306 & 3.48\\
    $Air$ & 1,226 & 2,410 & 3.93 & 0.0032 & 5.92 & 0.068 & 1.88\\
	$Road$ & 2,642 & 3,303 & 2.50 & 0.0009 & 35.35 & 0.016 & 1.09\\
	$Twitter$ & 3,212 & 3,423 & 2.13 & 0.0006 & 7.31 & 0.004 & 19.16 \\
    $Power$ & 4,941 & 6,594 & 2.67 & 0.0005 & 18.99 & 0.080 & 1.45\\
  \hline \hline
\end{tabular}}
\end{table}

\subsection{Artificial network models}
We have known that the BA and WS networks models are widely used to simulate real complex networks because they characterize real complex networks' stylized facts. Herein, we show two types of artificial network models that are used in the following research, which are described as:

\textbf{Barabasi-Albert network model}\cite{barabasi1999emergence}: The BA network model proposed by Barabasi and Albert characterizes the scale-free property of real complex networks. By using mean filed approximation, it can be proved that the resulted BA network has a power-law degree distribution with a scaling exponent 3. In the simulating process, the number of nodes and edges are adjustable according to the actual need.

\textbf{Watts-Strogatz network model}\cite{watts1998collective}: The WS network model proposed by Watts and Strogatz characterizes the small-world property of real complex networks. The resulted WS network has a larger cluster coefficient and shorter average distance. However, its degree distribution is Poisson. In the simulating process, the number of nodes and edges and the rewired probability are adjustable according to the actual need.



\subsection{Evaluation}
The common measuring index for evaluating the link prediction method is AUC, which refers to the area under the receiver operating characteristic curve (ROC) \cite{hanley1982meaning}. In the AUC calculation, we needn't draw the specific ROC curve, especially when the samples are enormous. Rather than, we generally use the sampling method to obtain its approximate value. Once the partition of the training set and the testing set is determined, there are two kinds of unknown edges in the training set. One corresponds to the nonexistent edges (i.e., they don't exist in both training and testing sets). The other corresponds to the hidden edges (i.e., they only exist in the testing set). For a given link prediction method, each unknown edge is given a similarity index. AUC is equivalent to the probability that the similarity index of the randomly selected hidden edge in the testing set is higher than that of randomly chosen nonexistent edges \cite{fawcett2006introduction}.

So, we randomly select a hidden edge and a nonexistent edge in the testing set. If the similarity index of the hidden edge is higher than that of the nonexistent edge, the AUC value is added by $1$. If these two similarity indexes are equal, the AUC value is added by $0.5$. The sampling process is repeated with $N$ times. We assume that there are $N^{\prime}$ and $N^{\prime\prime}$ times of the sampling processes that meet the two cases mentioned above, respectively. The AUC value is calculated as
\begin {equation}
AUC=\frac{N^{\prime}+0.5 \cdot N^{\prime\prime}}{N}
\label{eq28}
\end {equation}
Note that a larger $N$ makes the higher confidence of the measurement of AUC in equation (\ref{eq28}). According to \cite{Lv2005}, when $N \geq 672400$, we can guarantee with 90\% confidence that the absolute error of AUC will not exceed one-thousandth no matter of the network size. Thus, we set $N=672400$ in the measurement of AUC.

\section{Experimental result and discussion}
This section presents the link prediction method's performance based on SSNE and compares the proposed method with other baselines. The 20$\%$ edges of the current network $G_t$ is hidden to obtain the previous network $G_{t-1}$. There are no isolated nodes in both $G_{t-1}$ and $G_t$. Furthermore, we explore the effectiveness of adjustable parameters in the proposed method according to the experimental results based on real networks and artificial networks. Finally, we summarize the optimal AUC values obtained from the proposed method and the mainstream methods based on six real networks and two types of artificial network models.

\subsection{Link prediction in real networks}
Herein, we first examine the link prediction method's performance based on SSNE and compare the proposed method with several mainstream methods based on structural similarity indexes and graph embeddings, such as CN, AA, RA, RWR, and DeepWalk. More other methods are shown in the following summary of the experimental result. AUC is used to evaluate the link prediction performance of these methods. The order $h$ and dimension $d$ are considered adjustable parameters, which regulate the link prediction method based on SSNE. Because the full dimension $n$ is different from each network, $d$ is dependent on $n$, \emph{i.e.}, $d=p \cdot n$ for $p \in (0,1) $. Note that $p$ is an alternative parameter of $d$ that indicates the proportion of dimension reduction to network size. \textbf{Figure 2} presents the performance comparison of different link prediction methods for six real networks. It suggests that except the Yeast, the link prediction method based on SSNE (short of $SSNE(h,p)$) behaves better than these mainstream methods.
\begin{figure}[ht]
\centering
\includegraphics[width=\linewidth]{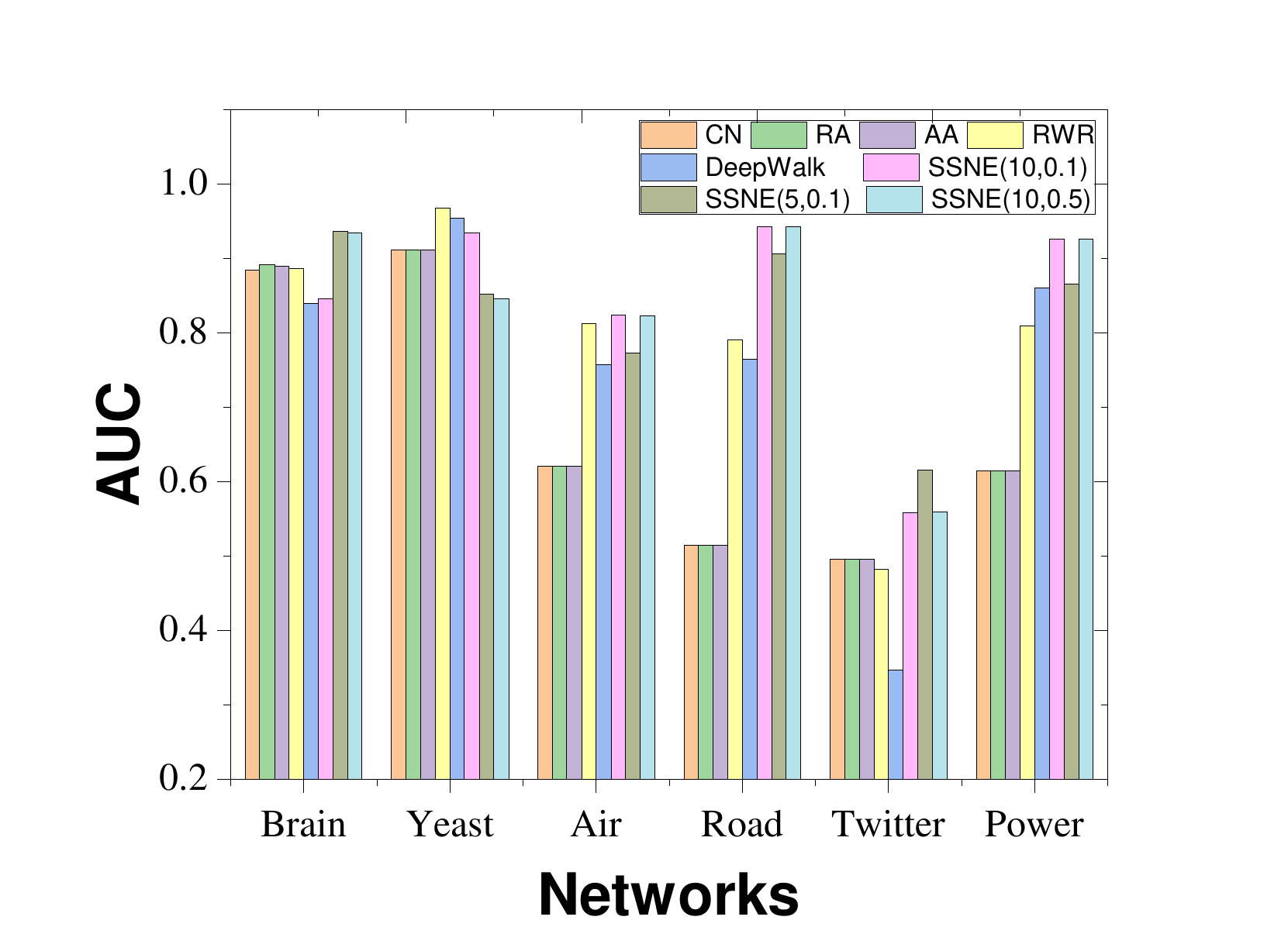}
\caption{(Color online) Performance comparison of different link prediction methods for six real networks. Except the Yeast, the link prediction method based on SSNE (short of $SSNE(h,p)$) behaves better than these mainstream methods.}
\end{figure}

More concretely, as shown in \textbf{Figure 2}, it is found that in these networks with the relatively large average degree (e.g., \textbf{Brain}, \textbf{Yeast}), the link prediction performance of the proposed method is similar to that of the method based on DeepWalk. Both of them do not significantly outperform other methods based on the structural similarity index. However, when the average degree is relatively small (e.g., \textbf{Road}, \textbf{Twitter}, \textbf{Power}), the proposed method performs the best. Thus, we think that the proposed method is more suitable to solve the link prediction problem of sparse networks.
Note that the artificial networks will further verify such observation in the following subsection.

\begin{figure*}[ht]
\centering
\includegraphics[width=\linewidth]{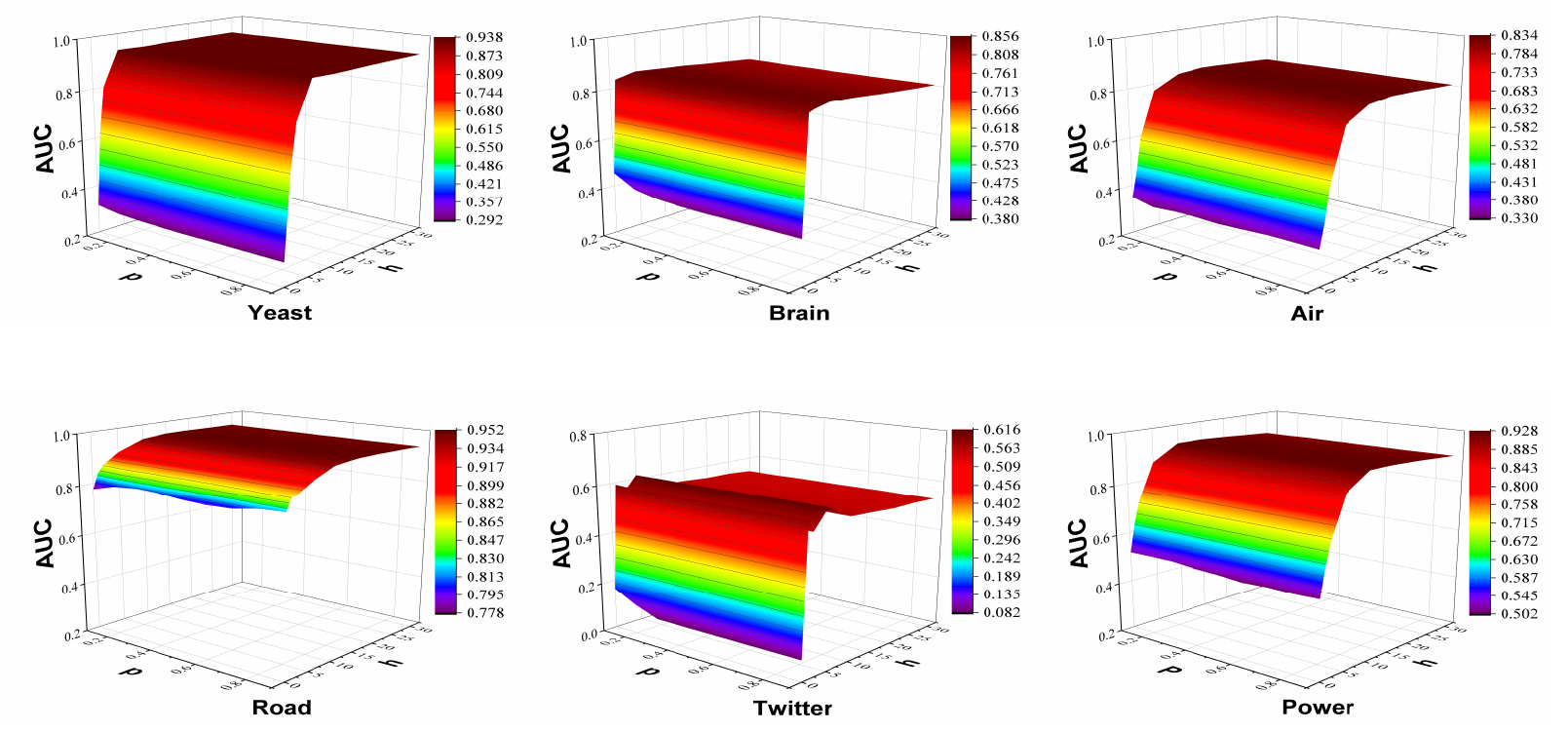}
\caption{(Color online) The Influence of both the order $h$ and the proportion $p$ on the link prediction method based on SSNE. For each network, we show the AUC values in respect to different combinations of $h$ and $p$. The results verify the stability of the proposed method because they share a similar trend in respect to $h$ and $p$.}
\end{figure*}

At the same time, it is also found that the proposed method is affected by the adjustable parameters. We use different combinations of order $h$ and proportion $p$ to comprehensively analyze the link prediction for six real networks. \textbf{Figure 3} presents the influence of both $h$ and $p$ on the link prediction performance based on six real networks. The best AUC values of six real networks are 0.938 of \textbf{Yeast}, 0.856 of \textbf{Brain}, 0.834 of \textbf{Air}, 0.952 of \textbf{Road}, 0.616 of \textbf{Twitter}, and 0.928 of \textbf{Power}. We find that the proposed method is not particularly sensitive to the changes of $h$ and $p$. More concretely, for a given $h$, the link prediction performance is nearly unchanged when $p$ varies from 0.1 to 0.9. We can easily understand that the operation of SVD in SSNE causes such a phenomenon. There exists a critical $d_c$ in each sparse network. The $d_c$-dimension feature matrix nicely represents the structural information. However, the acquirement of $d_c$ in each sparse network brings high computational costs. We use $p$ to uniformly set the corresponding dimensions of sparse networks for the simplicity of parameter computation. Even for $p=0.1$, the corresponding $d$-dimension feature matrices can well represent the complete structural information of these sparse networks. While for a given $p$, the link prediction performance changes primarily when $h$ gradually increases in a small range and then becomes approximately stable with the convergence of the SNHAM matrix, which implies that the SNHAM matrix with a small order (at least 10) contains most of the topological information of network structure. Furthermore, each network's results reveal a similar trend, which verifies the proposed method's stability. After the analysis mentioned above, we observe that when $h=10$ and $p=0.1$, the proposed method almost converges to the nearly optimal link prediction performance. It roughly suggests the default set of adjustable parameters can correspond to $h=10$ and $p=0.1$ for obtaining better link prediction performance.

\subsection{link prediction in artificial networks}
We also testify to the proposed method based on artificial networks. The artificial networks are generated by the BA and WS network models, respectively. Each type is composed of multiple artificial networks with various average degrees and sizes of nodes. Specifically, the sizes of nodes in the BA (or WS) networks vary from 1000 to 5000. For the BA (or WS) networks with the fixed size, their average degrees vary from 2 to 10 with a step-length two by adding edges which indicates the changes of edge sparsity. We try to study the relationship between the network sparsity and link prediction performance (i.e., AUC) obtained from the proposed method.

\begin{figure}[ht]
\centering
\includegraphics[width=\linewidth]{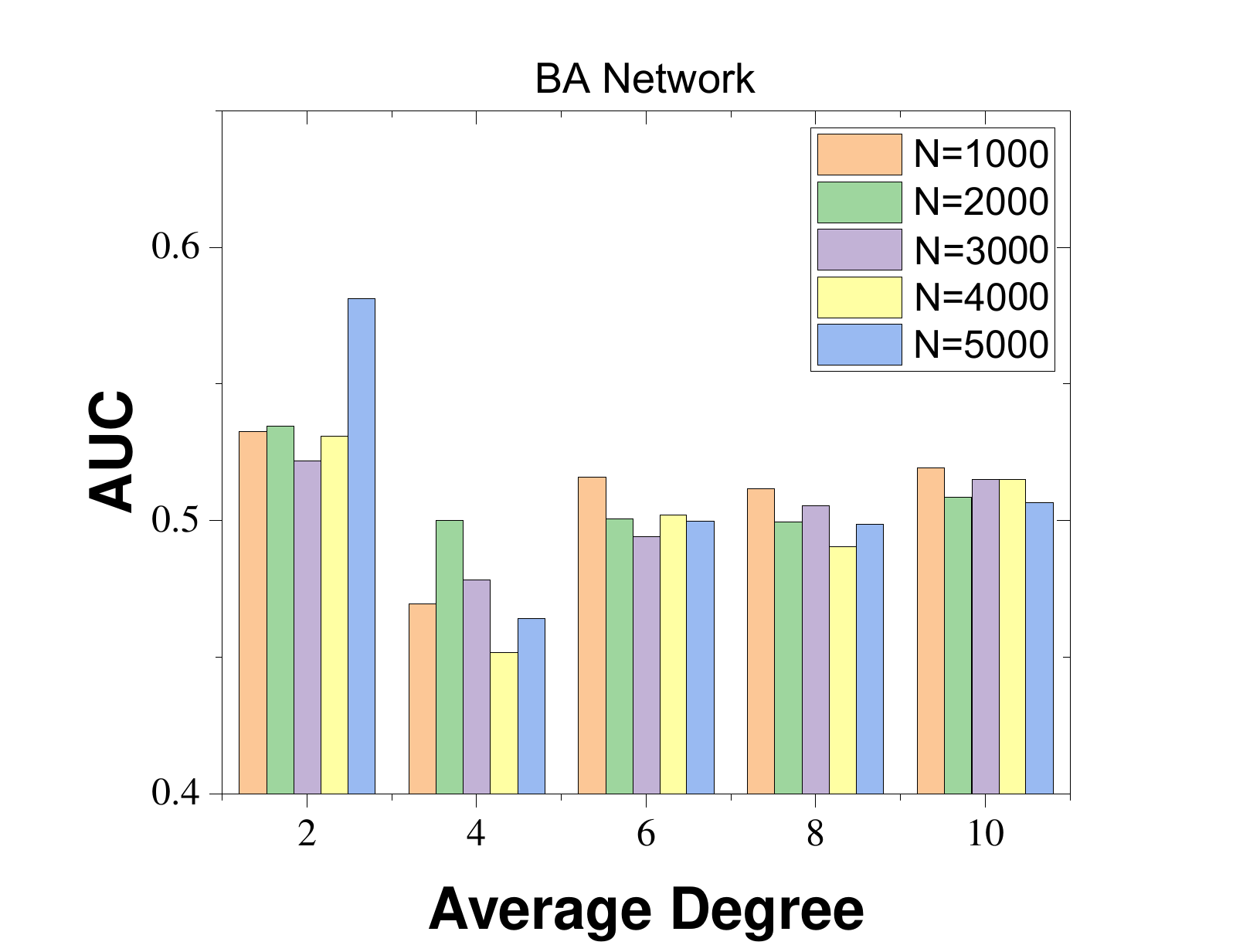}
\includegraphics[width=\linewidth]{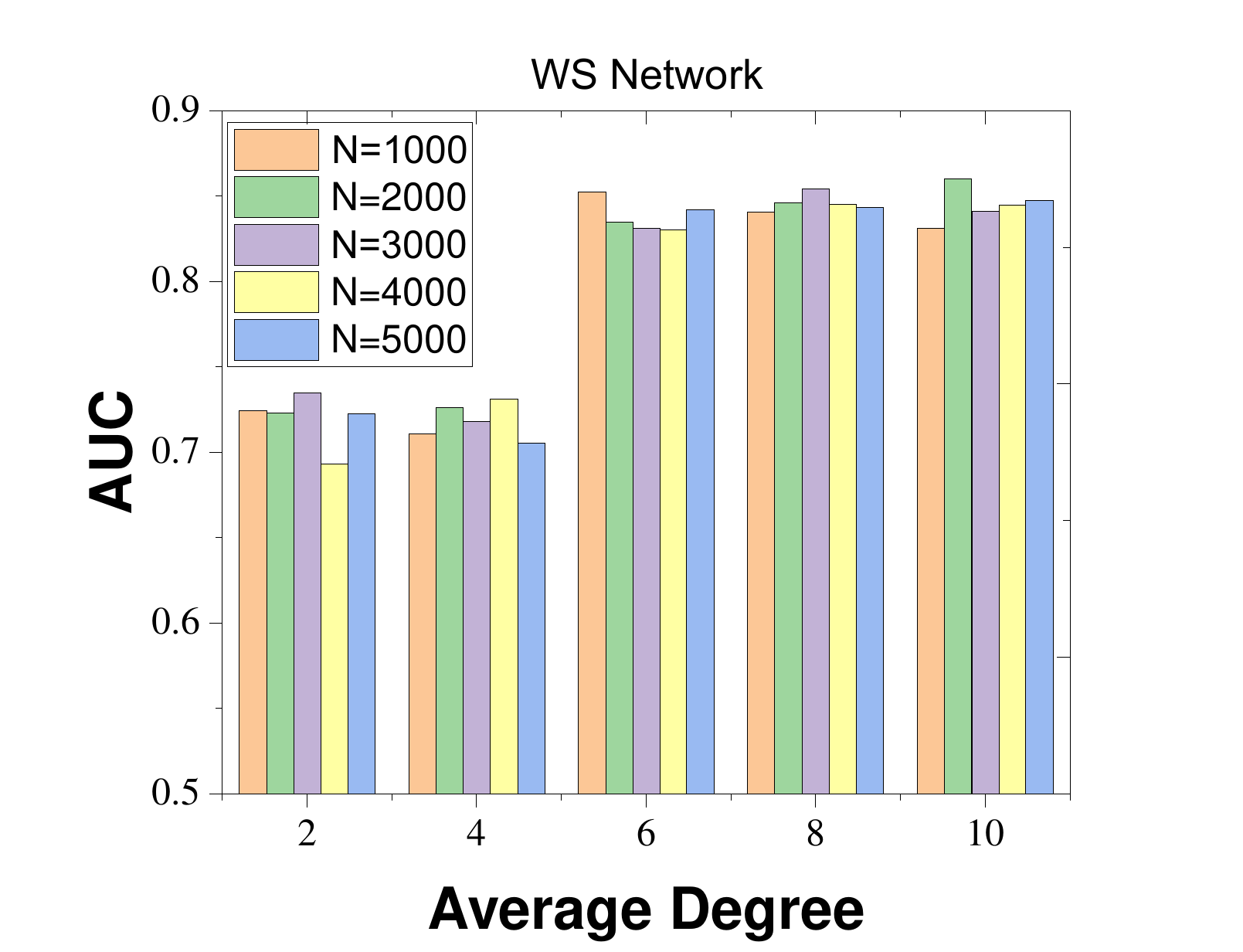}
\caption{(Color online) The link prediction performance of the proposed method based on the BA and WS networks with different average degrees and sizes of nodes.
In the upper panel, the AUC values as a function of the average degree show that the proposed method is much more suitable for the BA network with the relatively smaller average degree and lower edge sparsity. In the lower panel, the AUC values as a function of the average degree show that the result is contrary to that found in the BA networks. }
\end{figure}

\textbf{Figure 4} presents the AUC values as a function of the average degree, which are obtained from the link prediction in the BA and WS networks. As shown in the left panel of \textbf{Figure 4}, we can see that the link prediction performance is better realized by the proposed method when the BA networks have a relatively smaller average degree and lower edge sparsity (e.g., $\langle k \rangle = 2$ and $N=5000$). In particular, no matter the sizes of nodes, the AUC values are optimal when the average degrees of BA networks are $\langle k \rangle = 2$, which suggests that the proposed method is sensitive to the average degree. Meanwhile, as shown in the right panel of \textbf{Figure 4}, we can see that when the average degrees in the WS networks increase, the link prediction performance becomes much better, which is contrary to the results found in the BA networks. Nevertheless, when the average degree rises, the differences in link prediction performance between the proposed method and those based on structural similarity indexes become smaller (see in \textbf{Table 3}). In the sparse WS networks ($\langle k \rangle = 2$), the proposed method shows much better link prediction performance. Thus, to some extent, these results show that the proposed method is more suitable for link prediction in sparse networks.

\subsection{Discussion}
As we comprehensively analyze the proposed method's link prediction performance based on real networks and artificial networks, we further discuss the performance comparison between the proposed method and more mainstream methods by conducting extensive experiments. Note that the average degrees of artificial networks are set as 2 and 6, and their sizes are both 5000 nodes.

\textbf{Table 3} presents the performance comparison of all link prediction methods by the AUC values. The mainstream methods are divided into three types: structural similarity indexes including CN, Salton, Jaccard, AA, RA, RWR LHN-I, LHN-II, Katz SimRank, matrix optimization including CLMC \cite{zhang2019semisync}, and graph embedding models including DeepWalk (DW), Diff2Vec (D2V) \cite{rozemberczki2018fast}, Struc2Vec (S2V) \cite{ribeiro2017struc2vec}. The specific parameter sets are illustrated:
RWR with parameter $c=0.8$; SimRank with parameter $\lambda=0.8$; CLMC with parameters $\alpha_{1}=0.001$, $\alpha_{2}=0.01$ and $\alpha_{3}=100$; DeepWalk with parameters $Windows =10$, $length=40$, $times=30$, $d=128$; Diff2Vec with parameters $\alpha =0.025$, $Windows =10$, $vertex set cardinality=40$, $num diffusions =10$, $d=128$; Struc2Vec with parameters $times =20$, $Windows =5$, $length=40$, $d=64$; SSNE with default parameters $h=10$, $rate=0.1$.

More concretely, in \textbf{Table 3}, the first and second-best AUC values are underlined by bold characters. We can find that for these networks with the relatively large average degree (e.g., \textbf{Brain} and \textbf{Yeast} the link prediction performance obtained by structural similarity indexes is better than the other two types of link prediction methods. However, except for the second-best AUC values in the Air network, we can see that the proposed method (i.e., SSNE) achieves the best AUC values for the link prediction in these real networks relatively small average degree. In artificial networks, we can see that the proposed method performs the best AUC values for the link prediction in these sparse artificial networks (i.e., $\langle k \rangle =2 $).

\begin{table*}
  \caption{The performance comparison of link prediction methods. The first and second-best AUC values are underlined by bold characters. }
  \label{Table:3}
  \centering
  \resizebox{\textwidth}{!}{
  \begin{tabular}{c|c|c|c|c|c|c|c|c|c|c|c|c|c|c|c}
    \hline\hline
    Nets & CN & Salton &Jaccard &AA &RA &RWR &LHN-I &LHN-II &Katz &SimRank &CLMC &DW &D2V &S2V &SSNE \\
    \hline
    $Brain$ & 0.884 & 0.885 & \underline{\textbf{0.905}} & \underline{\textbf{0.891}} & 0.889 & 0.886 & 0.771 & 0.651 & 0.886 & 0.754 & 0.784 & 0.839 &0.793 &0.652 & 0.846  \\
    $Yeast$ & 0.911 & 0.910 & 0.913 & 0.912 & 0.912 & \underline{\textbf{0.967}} & 0.905 & 0.962 & \underline{\textbf{0.967}} & 0.674 & 0.928 & 0.954 
&0.947&0.271& 0.934 \\
    $Air$ & 0.620 & 0.619 & 0.626 & 0.620 & 0.620 & 0.813 & 0.619 & 0.781 & 0.813 & 0.674 & 0.621 & 0.757 &\underline{\textbf{0.863}} &0.350& \underline{\textbf{0.823}} \\
	$Road$ & 0.514 & 0.514 & 0.515 & 0.514 & 0.514 & 0.791 & 0.514 & 0.779 & 0.779 &0.931& 0.202 & 0.764 &\underline{\textbf{0.940}} &0.529& \underline{\textbf{0.942}} \\
	$Twitter$ & 0.496 & 0.496 & \underline{\textbf{0.507}} & 0.496 & 0.496 & 0.483 & 0.496 & 0.465 & 0.483 & 0.432 & 0.479 & 0.346 &0.469 &0.135& \underline{\textbf{0.558}} \\
    $Power$ & 0.614 & 0.614 & 0.614 & 0.614 & 0.614 & 0.809  & 0.614 & 0.809 & 0.809 & \underline{\textbf{0.919}} & 0.505 & 0.860 &0.907 &0.450& \underline{\textbf{0.927}}  \\
	$BA(2)$ & 0.499 & 0.499 & \underline{\textbf{0.500}} & 0.499 & 0.499 & 0.477 & 0.499 & 0.477 & 0.477 & 0.346 & 0.354 & 0.374 &0.370 &0.181& \underline{\textbf{0.581}} \\
	$BA(6)$ & 0.519 & 0.518 & 0.526 & 0.519 & 0.519 & 0.599 & 0.518 & 0.497 & 0.599 & \underline{\textbf{0.646}} & 0.507 &0.582 &\underline{\textbf{0.640}} &0.242& 0.499 \\
	$WS(2)$ & 0.501 & 0.501 & 0.501 & 0.501 & 0.501 & 0.498 & 0.501 & 0.498 & 0.498 & 0.551 & 0.516 &0.645 &\underline{\textbf{0.693}} &0.561& \underline{\textbf{0.713}} \\
	$WS(6)$ & 0.769 & 0.770 & 0.771 & 0.770 & 0.770 & 0.826 & 0.770 & 0.828 & 0.826 & 0.306 & 0.753 & 0.832 &\underline{\textbf{0.835}} &0.482& \underline{\textbf{0.843}} \\
  \hline\hline
\end{tabular}}
\end{table*}

Finally, we quantitatively analyze each link prediction method's running efficiency via a personal computer with the 20 Intel(R) Xeon(R) CPU and 64G DDR4 memory. Note that the running time is directly presented to supplement experimental analysis. \textbf{Table \ref{Table:4}} shows the running time of link prediction methods based on artificial and real sparse networks. It can be seen that the highest running time is $3294.3$ seconds of SimRank, which suggests that realizing the link prediction in both artificial and real sparse networks is feasible. Besides, we discuss the running efficiency of link prediction methods. For these link prediction methods (except SimRank) based on structural similarity indexes, the running time is relatively stable and much less than that of link prediction methods based on matrix optimization and graph embedding models. The high running efficiency is because the running time mostly spends in the computation of structural similarity indexes. For CLMC and SSNE, the low running efficiency is because the running time mostly spends in the multiple iterations of matrix computation (e.g., the matrix computation in SNHAM algorithm). Note that the running time of DeepWalk, Diff2Vec, and Struc2Vec is much less because the pre-training time of node representation is neglected.

 \begin{table*}
  \caption{The running time (in seconds) of link prediction methods.}
  \label{Table:4}
  \centering
  \resizebox{\textwidth}{!}{
  \begin{tabular}{c|c|c|c|c|c|c|c|c|c|c|c|c|c|c|c}
    \hline\hline
    Nets & CN & Salton &Jaccard &AA &RA &RWR &LHN-I &LHN-II &Katz &SimRank &CLMC &DW &D2V &S2V &SSNE \\
    \hline
    $Brain$ & 105.0 & 102.0 & 97.0 & 107.2& 106.9 & 0.1 & 0.1 & 0.1 & 0.1 & 20.8 & 1.3 & 23.5 &14.1 &16.6 & 16.0  \\
    $Yeast$ & 100.2 & 100.5 & 103.5 & 102.0 & 101.9 & 0.8 & 0.4 & 1.0 & 0.6 & 910.4 & 245.4 & 132.4 &18.9&22.3& 214.9 \\
    $Air$ & 87.6 & 87.3 & 85.2 & 89.1 & 88.6 & 0.3 & 0.2 & 0.3 & 0.2 & 74.9 & 34.8 & 73.0 & 15.2 &17.4& 68.3 \\
    $Road$ & 82.0 & 83.5 & 84.5 & 81.9 & 81.5 & 1.5 & 0.5 & 1.3 & 1.2 & 252.8 & 361.2 & 126.0 & 21.2 & 24.3 & 262.0 \\
    $Twitter$ & 84.2 & 83.7& 85.7 & 86.1 & 84.4 & 1.6 & 0.7 & 2.0 & 1.2 & 356.7 & 593.9 & 141.1 &24.1 & 27.8& 384.8 \\
    $Power$ & 84.9& 84.9 & 85.8 & 86.2 & 85.5 & 5.1 & 1.4  & 5.5 & 3.5 & 951.5 & 1832.4 & 237.8 & 38.8 &42.6 &889.8  \\
    $BA(2)$ & 86.1 & 85.5 & 88.7 & 88.5 & 87.3 & 6.5 & 2.2 & 8.0 & 5.2 & 1281.4 & 2620.8 & 274.8 &51.0 &59.3& 1312.8 \\
    $BA(6)$ & 97.4 & 90.9 & 96.2 & 97.3 & 97.5 & 7.2 & 2.2 & 8.5 & 5.3 & 2437.2 & 2585.0 &333.4 &52.5 &56.8& 1329.2 \\
    $WS(2)$ & 81.9 & 83.6 & 80.8 & 83.0 & 81.4 & 6.5 & 2.5 & 8.3 & 5.3 & 1254.3 &2624.3 &254.4 &50.5& 57.4 & 1341.8 \\
    $WS(6)$ & 87.1 & 93.1 & 88.5 & 88.8 & 87.4 & 5.7 & 2.1 & 7.5 & 4.8 & 3294.3 & 2274.6 & 340.8 & 49.2&56.3& 1328.2 \\
  \hline\hline
\end{tabular}}
\end{table*}

\section{Conclusion}

As graph embedding is recently used for link prediction in complex networks, this paper proposes a novel link prediction method based on SSNE constructed in the framework of graph embedding. We comprehensively describe the procedure of SSNE from two aspects, the SNHAM matrix and the neural network model. The SNHAM matrix contains the $h$-order structural information of the adjacency matrix, and the neural network model is used to learn the $d$-dimensional representation of the SNHAM matrix. Through the SSNE, we can effectively obtain the graph representation of network structure.
Note that the graph embedding procedure of SSNE is irrelevant to a specific network structure. Most importantly, in the SSNE, the adjustable parameters have been significantly reduced into two variables. Thus, the SSNE overcomes the random-walk-based graph embedding models' critical drawbacks by avoiding a directly random walk sampling process.

Meanwhile, to reduce the computational complexity of the neural network model, we assume that the optimization procedure of mining the loss function is equivalent to making the output matrix approximate the SNHAM matrix by adjusting the kernels of the neural network model. The product matrix of the kernels denotes the output matrix. Then, we formalize the association between the SNHAM matrix and the product matrix through the Softmax function. By verifying the inverse softmax function's assumption, we obtain the product matrix
indicated by the logarithmic SNHAM matrix. Finally, we use the SVD to solve the product matrix and get the $d$-dimensional feature matrix.

The link prediction method based on the feature matrix is constructed by calculating the similarity indexes among feature vectors. We use six real networks and two types of artificial network models to test the proposed method's link prediction performance. The testing experiments are designed in three aspects. We first verify the proposed method's effectiveness on the link prediction in real diverse networks and the adjustable parameters' sensitivity to the proposed method. It has been found that the proposed method is more suitable for the link prediction in the relatively sparse network and only partially sensitive to the order of the SNHAM matrix. Then, the proposed method's effectiveness on the link prediction in the sparse network is further verified based on artificial networks. Finally, we discuss the comparison of the proposed method with a lot of mainstream methods based on structural similarity indexes, matrix optimization, and other graph embedding models. It suggests that the proposed method shows better link prediction performance in a relatively sparse network.

\bibliographystyle{IEEEtran}
\bibliography{reference}

\begin{IEEEbiography}[{\includegraphics[width=1in,height=1.25in,clip,keepaspectratio]{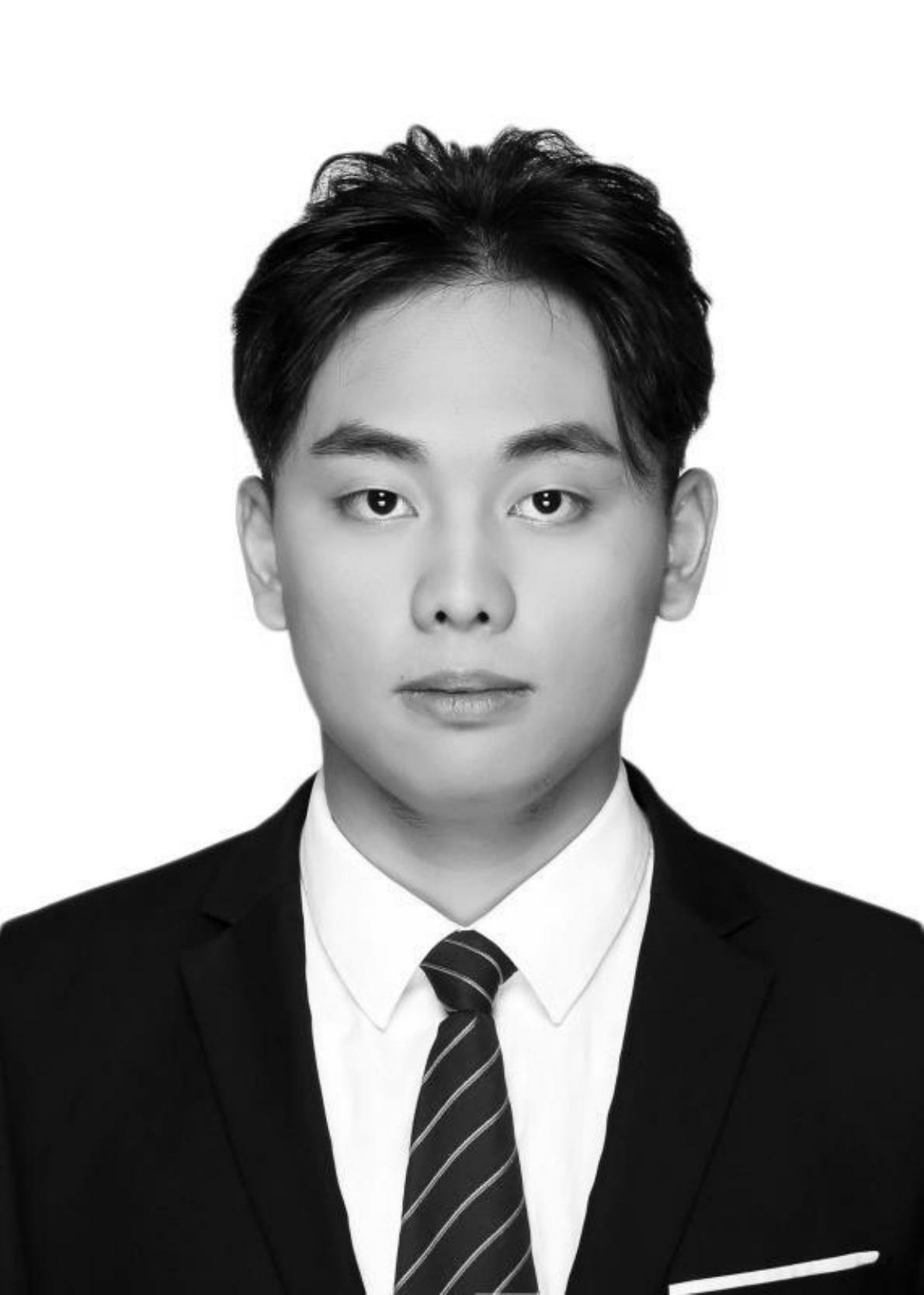}}]{MING-REN CHEN} received the B.S. degree in information security from University of Electronic Science and Technology of China, Chengdu, China, in 2019. He is currently pursuing the M.S. degree in computer science and technology with the University of Electronic Science and Technology of China, Chengdu, China, since 2019, and has been studying in the Computer science and engineering, University of Electronic Science and Technology. His main awards and honors include Merit Student, the National Encouragement Scholarship and outstanding graduates. His main interests include data mining, deep learning and complex network.
\end{IEEEbiography}

\begin{IEEEbiography}[{\includegraphics[width=1in,height=1.25in,clip,keepaspectratio]{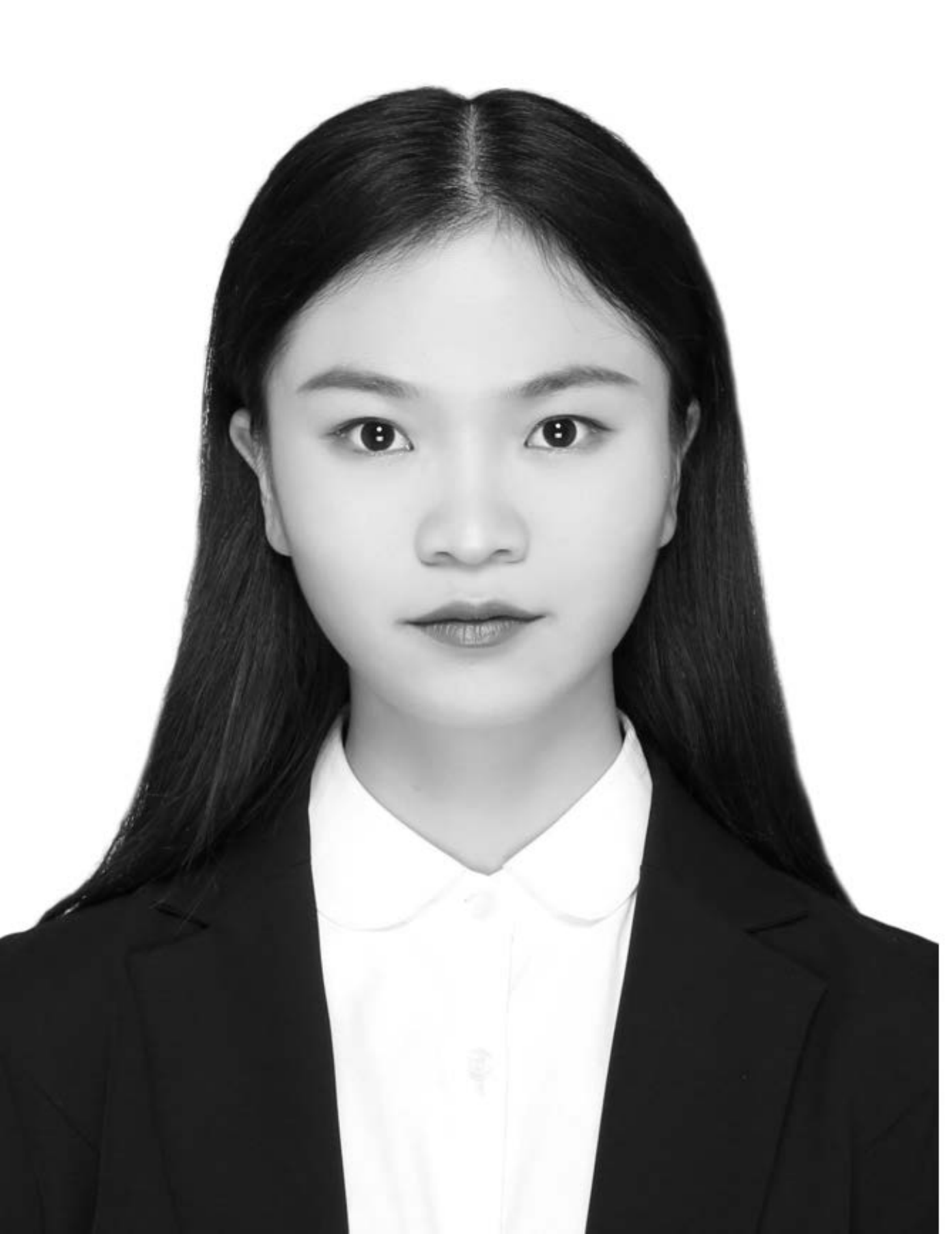}}]{PING HUANG} received the B.S. degree in electronic science and technology from Southwest Jiaotong University, Chengdu, China, in 2018. She is currently pursuing the M.S. degree in computer science and technology with the University of Electronic Science and Technology of China, Chengdu, China, since 2018, and has been studying in the Computer science and engineering, University of Electronic Science and Technology. Her main awards and honors include Merit Student, the first prize scholarship and outstanding graduates.Her main interests include complex network, information diffusion and epidemic spreading.	
\end{IEEEbiography}

\begin{IEEEbiography}[{\includegraphics[width=1in,height=1.25in,clip,keepaspectratio]{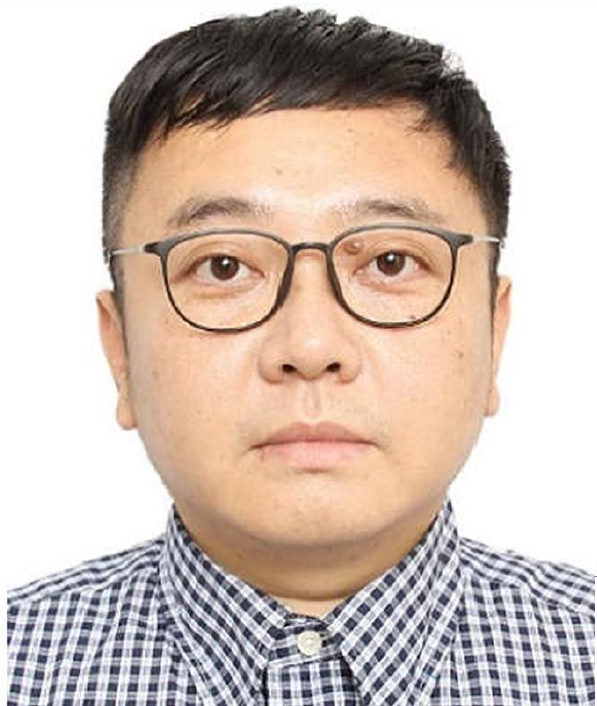}}]{YU LIN} received the Ph.D. degree in computer science from EPFL, Lausanne, Switzerland. He was a postdoctoral scholar at the Department of Computer Science and Engineering, University of California, San Diego. Currently, he is a Lecturer at the Research School of Computer Science, Australian National University. His research interests include algorithm design and computational biology.
\end{IEEEbiography}

\begin{IEEEbiography}[{\includegraphics[width=1in,height=1.25in,clip,keepaspectratio]{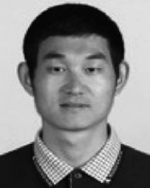}}]{SHI-MIN CAI} received the B.S. degree in electrical engineering and automation from the Hefei University of Technology, in 2004, and the Ph.D. degree in circuit and systems from the University of Science and Technology of China, in 2009. He currently serves as an Associate Professor of the University of Electronic Science and Technology of China. At present, he has published more than 100 high-level academic articles and hosted/participated 9 national projects mainly supported by the National Natural Science Foundation of China. He is interested in complex network theory and its application for mining and modeling of real large-scale networked systems, time series analysis, and personalized recommendation systems.
\end{IEEEbiography}

\EOD

\end{document}